\documentclass[runningheads]{llncs}
\usepackage[T1]{fontenc}

\usepackage{graphicx}
\usepackage{algorithm}
\usepackage{algorithmic}
\usepackage{multirow}
\usepackage{booktabs}
\usepackage{graphicx}
\usepackage{textcomp}
\usepackage{array}
\usepackage{cite}
\usepackage{amsmath,amssymb,amsfonts}
\usepackage{epstopdf}
\usepackage[table,xcdraw]{xcolor}
\usepackage{float} 
\usepackage{subfigure} 
\usepackage[justification=centering]{caption}

\begin{document}
\title{Feature-selected Graph Spatial Attention Network for Addictive Brain-Networks Identification}
\titlerunning{Feature-selected Graph Spatial Attention Network}
%
\author{Changwei Gong\inst{1} \and
Changhong Jing\inst{1} \and
Junren Pan\inst{1} \and
Shuqiang Wang\inst{1}}
\authorrunning{Changwei Gong et al.}
%
\institute{Shenzhen Institutes of Advanced Technology, Chinese Academy of Sciences,\inst{1} Shenzhen, China \\
\email{cw.gong@siat.ac.cn}\\
\email{ch.jing@siat.ac.cn}\\
\email{jr.pan@siat.ac.cn}\\
\email{sq.wang@siat.ac.cn}}
\maketitle              
\begin{abstract}
Functional alterations in the relevant neural circuits occur from drug addiction over a certain period. And these significant alterations are also revealed by analyzing fMRI. However, because of fMRI's high dimensionality and poor signal-to-noise ratio, it is challenging to encode efficient and robust brain regional embeddings for both graph-level identification and region-level biomarkers detection tasks between nicotine addiction (NA) and healthy control (HC) groups. In this work, we represent the fMRI of the rat brain as a graph with biological attributes and propose a novel feature-selected graph spatial attention network(FGSAN) to extract the biomarkers of addiction and identify from these brain networks. Specially, a graph spatial attention encoder is employed to capture the features of spatiotemporal brain networks with spatial information. The method simultaneously adopts a Bayesian feature selection strategy to optimize the model and improve classification task by constraining features. Experiments on an addiction-related neural imaging dataset show that the proposed model can obtain superior performance and detect interpretable biomarkers associated with addiction-relevant neural circuits.

\keywords{Neural imaging computing  \and Brain networks \and Graph attention \and Generative deep learning}
\end{abstract}
\section{Introduction}
Medical image computing is becoming increasingly significant in diseases analysis \cite{wang2015pre, pan2021decgan, zuo2021multimodal,wang2018automatic} because it enables the extraction and use of quantitative picture characteristics from routine medical imaging at high throughput, hence improving diagnostic, prognostic, and predictive accuracy. Neuroimaging, a bridge field integrating medical imaging computers and neuroscience, has also grown in recent years. Neuroimaging is a powerful tool for studying neuroscience and interpreting the anatomical structure and activity of the brain through qualitative and quantitative analysis of images in two and three dimensions by using imaging methods, as well as for resolving unresolved issues in the field of neurosciences and diagnosing and treating brain diseases.

The development of MRI changed the study of neuroanatomy by making it feasible to conduct in vivo experiments with sufficient contrast to different brain regions for the first time. Researchers may estimate the microstructure inside a voxel using a computational method for MRI-based neuroanatomical investigations. The capacity to gather high-quality, comprehensive information from in vivo imaging has certainly shaped researchers' knowledge of neuroanatomical and structure-function interactions and provided new insights into various disease processes. The introduction of functional neuroimaging in the past three decades has raised much enthusiasm about its potential to revolutionize researchers' knowledge of the physical foundation of the brain and to offer valuable tools for clinical and research study. Functional magnetic resonance imaging (fMRI) and resting-state functional magnetic resonance imaging (rs-fMRI)\cite{ref_1} are the most powerful noninvasive functional imaging techniques available at the time. Functional connection\cite{ref_3} in brain networks is often generated through the observation of fMRI time series, and functional brain networks\cite{ref_2} describe statistical correlation patterns between neuronal regions. Significant progress has been made in functional brain network analysis using fMRI data over the last decade. The variation of functional connectivity between brain areas has been widely investigated in terms of brain diseases\cite{ref_6}, as well as the association between cognitive impairments and degenerative neurological and psychiatric disorders\cite{02wang2020ensemble}.

Addiction is a disorder of the functioning brain characterized by abnormal behavior\cite{ref_4}. Addicts are driven by an overwhelming need to seek and consume drugs constantly. Although after prolonged withdrawal and awareness of the harmful health implications of drug use and the detrimental impact on family and society, addicts face the risk of relapse. Pioneering functional MRI studies have shown that nicotine activates various brain regions\cite{ref_5}. However, few neuroimaging computational approaches have used functional MRI to investigate the relationship between nicotine addiction and altered neuronal activity patterns throughout the brain and identify these patterns and detect regional neuroimaging biomarkers. Therefore, research into the neural mechanisms and supporting diagnoses associated with nicotine and other drug addiction has become increasingly critical.

\subsubsection{Related work}Machine learning technology has been widely used in the recognition of natural scenes\cite{ref_00}\cite{wang2018classification} \cite{wang2018skeletal}. In brain image computing, machine learning-based artificial intelligence technology has a broader scenario to make the analysis technology of brain images sink, effectively improving the efficiency and diagnostic accuracy of physicians' treatment\cite{03yu2020multi}. Brain image analysis technology for brain disease research can explore the disease's mechanism and understand the brain's operation process\cite{04you2022fine}. Recent advances in machine learning, particularly in deep learning, are helping to identify, classify and quantify existing brain images\cite{05yu2021tensorizing,pan2021char}. At the heart of these advances is the ability to automatically generalize hierarchical features from data rather than manually discovering and designing features based on domain-specific knowledge, as was previously the case\cite{06hu2020brain}. Deep learning is rapidly becoming state-of-the-art and replacing many original machine learning-based algorithms. Due to the development of deep learning, various neuroscience applications have simultaneously improved their performance significantly. Deep learning methods are a new approach for processing high-dimensional brain image data and extracting low-dimensional features\cite{07hu2021bidirectional} \cite{hu2019cross}\cite{hu2020medical}. For instance, convolutional neural network (CNN) methods\cite{01wang2018bone} reduce the dimensionality of medical image data by convolution operator to allow the effective identification of patterns in neuroimaging.
Generative adversarial strategies can simulate the real distribution of data to reduce the interference caused by noise and enhance the robustness of the model\cite{08yu2021morphological}. Generative Adversarial Network (GAN), which is based on variational methods\cite{09wang2009variational,wang2007variational,10mo2009variational,wang2008var}, is widely used in medical image computing \cite{wang2020diabetic}. GAN is usually unsupervised in training, and the newly generated data have the same distribution as the real data, thus allowing robust and complex representation learning, making it a commonly used method in medical image computation. However, existing methods for processing network structured data to obtain interpretable and deterministic biological markers are still challenging, especially for high-dimensional and sparse network datasets.

To solve these issues, we developed a novel feature-selected graph spatial attention network(FGSAN) to identify the efficient patterns of addiction-relevant brain networks from fMRI data and detect discriminative biomarkers of addiction-relevant brain regions that are interpreted by neuroscientific addiction circuit mechanism. The designed feature selection is efficient for graph representation learning and obtaining more helpful brain network embeddings to extract more accurately addiction-relevant biomarkers.

\begin{figure}
	\includegraphics[width=\textwidth]{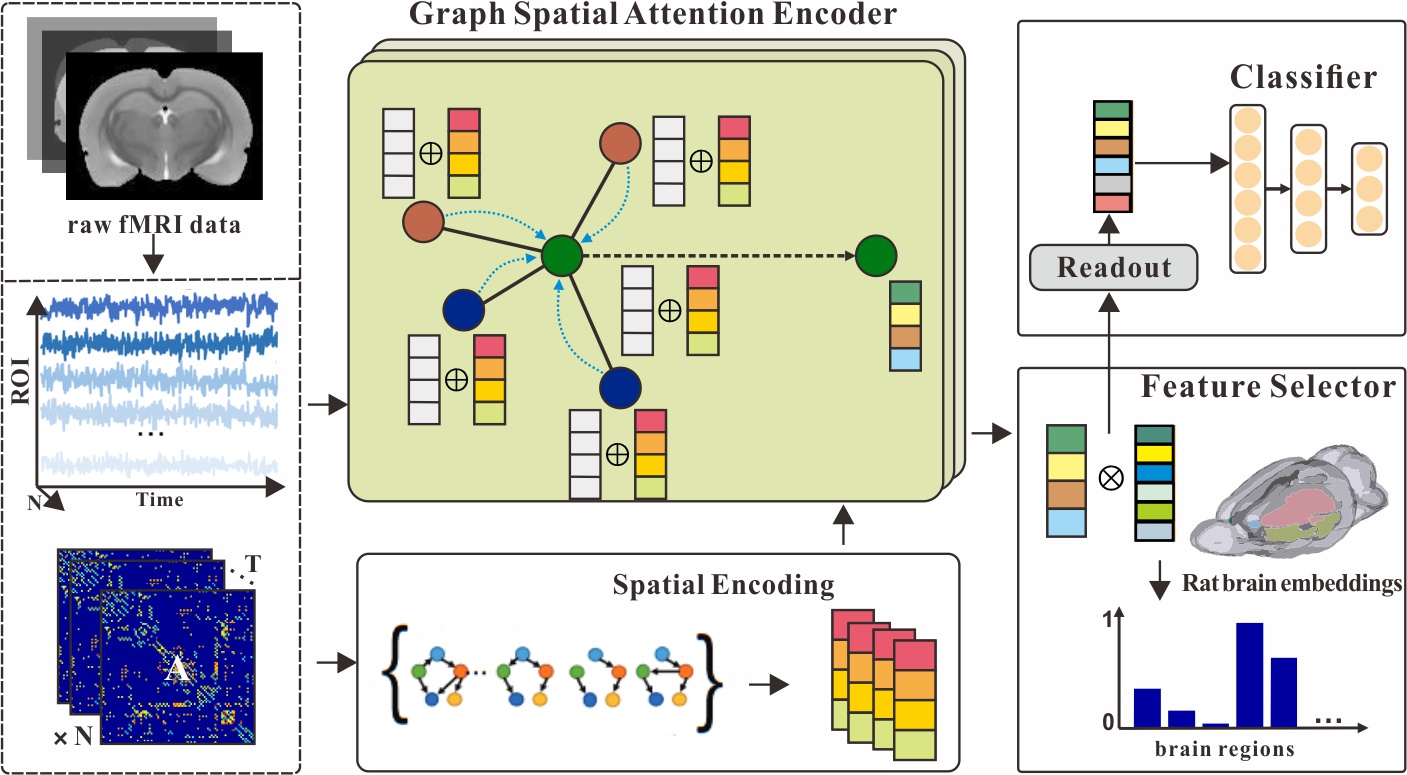}
	\caption{Proposed feature-selected graph spatial attention network for identifying brain addiction.} \label{fig1}
\end{figure}

\section{Method}
Our FGSAN is composed of three primary components: 1) an encoder consists of graph positional attention layer; 2) a feature selector with bayesian feature selection strategy; and 3) a classifier. The specific architecture are demonstrated in Fig.~\ref{fig1}.

In the encoder($E$), self-attention mechanism is adopted to transform the time series of brain regions $X=\left\{x_n\right\}^N_{n=1}\in\mathbb{R}^{N\times D}$  and dynamic brain functional connectivity $\left\{A^t\right\}^T_{t=1}$ into the embeddings $Z=\left\{z_n\right\}^N_{n=1}\in\mathbb{R}^{N\times d}$. Moreover, in the feature selector, latent binary random vectors $B=\{b_n\}^N_{n=1}$ are introduced to infer the posterior probability distribution and select more efficient brain regional features.  Therefore, the encoder is trained with double objectives: a bayesian feature selection loss considered as the feature sparsity penalty, and a classification loss  for identifying nicotine addiction.

\subsection{Graph Spatial Attention Encoder}
The graph spatial attention encoder aims to embed the regional brain imaging features aggregated with dynamic brain network attributes into a low-dimensional latent space. The proposed layer that composes the encoder is based on the graph attention networks(GAT)\cite{D1velivckovic2017graph} with the addition of spatial encoding. It allows each regional brain node to focus adaptively on other nodes according to the spatial information of the graph-structure connectivities in the brain networks.

Therefore, the attention coefficient, which is combined a shared attentional mechanism and spatial encoding for brain connectivities, can be expressed as:
\begin{equation}
	\mathbf{\alpha}_{n}^{l}(i, j)=\frac{\exp \left(\tanh \left(\left[\mathbf{h}_{n}^{l}(i) \mathbf{W}^{l} \quad \mathbf{h}_{n}^{l}(j) \mathbf{W}^{l}\right] \cdot \mathbf{c}^{l}+s_{\psi(x_i,x_j)}\right)\right)}{\sum_{j \in \mathcal{N}(i)} \exp \left(\tanh \left(\left[\mathbf{h}_{n}^{l}(i) \mathbf{W}^{l} \quad \mathbf{h}_{n}^{l}(j) \mathbf{W}^{l}\right] \cdot \mathbf{c}^{l}+s_{\psi(x_i,x_j)}\right)\right)}
\end{equation}
here $h^l_n(i)$ is a hidden representation for brain node $i$ at $l$th layer, $W^l\in\mathbb{R}^{d_l\times d_{l+1}}$ is a parameterized weight matrix considered as the graph convolutional filter,$c^l$ is a weight vector that can be learned in the train phase, and $S_{\psi(x_i,x_j)}$ is a scalar that can be learned and is indexed by $\psi(x_i,x_j)$ with positional information. It indicates the spatial encoding and is accessible throughout all layers.

Formally, let $\mathbf{h}_{n}^{l+1}(i)$ represent the output representation at $l$th layer, our graph spatial attention layer is given as follow:
\begin{equation}\label{eq2}
	\mathbf{h}_{n}^{l+1}(i)=\sigma\left(\sum_{j \in \mathcal{N}(i)} \mathbf{\alpha}_{n}^{l}(i, j) \mathbf{h}_{n}^{l}(j) \mathbf{W}^{l}\right)
\end{equation}
In Eq.\ref{eq2}, the feature propagation mechanism aggregates the effects across overall neighboring brain nodes and attaches spatial encoding information from dynamic brain network connectivity $\left\{A^t\right\}^T_{t=1}$.
\subsection{Bayesian Feature Selector}
To find the most effective features for identification from many regional brain features and to acquire a set of fewer but discriminative biomarkers to reduce classification error, we employ the bayesian feature selector. We define $\mathbf{H}=\{H_1^o,...,H_n^o\}$ and $\mathbf{Y}=\{y_1,...,y_n\}$ as the output features from the encoder and labels of addiction or not. By introducing binary masking matrix $B$ to achieve the goal of selecting features, the expected posterior distribution is denoted as $p(\mathbf{B}\mid \mathbf{H},\mathbf{Y})$ and an approximate distribution is represented as $q(\cdot)$. In order to improve the identification performance and the accuracy of the model in discriminating features, in the view of Bayesian inference, we optimize the model by minimizing the KL divergence between the posterior distribution and the approximate distribution:

\begin{equation}\label{eq3}
	\underset{q(\cdot)}{\operatorname{argmin}}KL(q(\mathbf{B})\|p(\mathbf{B}\mid \mathbf{H},\mathbf{Y}))=-E_{q}\left[\log \left(p(\mathbf{B}\mid \mathbf{H},\mathbf{Y})\right)\right]+K L\left(q\left(\mathbf{B}\right) \| p\left(\mathbf{B}\right)\right)
\end{equation}
In Eq.\ref{eq3}, the first term corresponds to a binary cross entropy loss for identification task where the input features $\mathbf{H}$ are masked by $\mathbf{B}$, and the second term becomes a loss for learning the probability scores $\mathbf{z}$ which is used to compute the binary matrix $\mathbf{B}$ by Bernoulli sampling method:
\begin{equation}
	\mathbf{b}_{n}=\sigma\left(\frac{\log \left(\mathbf{z}\right)-\log \left(1-\mathbf{z}\right)+\log \left(\mathbf{u}_{n}\right)-\log \left(1-\mathbf{u}_{n}\right)}{r}\right)
\end{equation}
where $\mathbf{u}_n$ is sampled from a uniform distribution from 0 to 1, and $r$ is the relaxation parameter of Bernoulli sampling.

\subsection{Classifier and Loss Function}
To integrate the information of each node for the graph-level identification, we utilize a readout function to cluster  node features together by simply averaging them:
\begin{equation}
	\mathcal{R}(\mathbf{H})=\sigma\left(\frac{1}{N} \sum_{i=1}^{N} \vec{h}_{i}\right)
\end{equation}
where $\sigma$ is nonlinear activation function. The readout function is similar to the graph pooling operation. Other graph pooling methods can be used to replace it. The selected and readout features are delivered to a multi-layer perceptron(MLP) to derive the final identification of predicted labels $\hat{y}$.

The total loss function is the interpretation of Eq.\ref{eq3}:
\begin{equation}
	\mathcal{L}\left(\mathbf{X}, \mathbf{A}\right)=-\sum_{n=1}^{N}\left(y_{n} \log \left(\hat{y}_{n}\right)+\left(1-y_{n}\right) \log \left(1-\hat{y}_{n}\right)\right)+K L\left(\operatorname{Ber}\left(\mathbf{z}\right) \| \operatorname{Ber}(\mathbf{s})\right)
\end{equation}
The first term is used to guide the MLP in the classification of the selected features. Furthermore, the second term is applied for training the selector to learn the probability mapping to the feature mask. Here $\operatorname{Ber}(\mathbf{s})$ is a binary random vector that contains sparse elements for the purpose of complying with sparsity.

\section{Experiments}

\subsubsection{Dataset and Preparation.}
The animal addiction experiment dataset contains two types of data with equal numbers: functional MRI images of nicotine non-addicted and nicotine addicted, each with $800$ time series. By preprocessing long-term functional MRI scans of experimental rats, we were able to create the dynamic brain network dataset needed for the experiment. The Statistical Parametric Mapping 8 (SPM8) program was used to do the first preprocessing in MATLAB. Functional data were aligned and unwarped to account for head motion, and the mean motion-corrected picture was coregistered with the high-resolution anatomical T2 image. The functional data were then smoothed using an isotropic Gaussian kernel with a $3mm$ full-width at half-maximum (FWHM). $150$ functional network regions were identified using the Wister rat brain atlas. We assessed the spectral link between regional time series using magnitude-squared coherence, resulting in a $150\times150$ functional connection matrix for each time step, whose members represented the intensity of functional connectivity between all pairings of regions.
\subsubsection{Implementation Detail.}
The PyTorch backend was used to implement FGSAN. One Nvidia GeForce RTX 2080 Ti was used to speed up the network's training. During training, the learning rate was set at 0.001, and the training epoch was set to 500. Adam was used as an optimizer with a weight decay of 0.01 to reduce overfitting. We construct the encoder with three graph spatial attention layers. All trials are repeated ten times, and the results are averaged. The regularization value was set to 0.5 for all datasets and techniques.
\subsubsection{Metrics.}
Evaluation of binary classification performance is based on quantitative measures in four key metrics: 1) accuracy (ACC); 2) Precision (PREC); 3) Sensitive (SEN); and 4) Specificity(SPEC). Our proposed method is evaluated by 8-fold cross-validation.

\subsection{Ablation Study}

As indicated in Table \ref{tab1}, we conducted ablation research on identification to evaluate the effectiveness of our proposed encoder and bayesian feature selector, and two significant results were achieved: 1) In the comparison of the baseline encoder approach showed impressive performance on the binary addiction-related classification. This is due to the fact that the attachment of spatial encoding enables the attention mechanism to get more positional information and learn better graph representations;  2) The approach with feature selector is generally performed well. It represents that feature selection plays its role as an auxiliary to identifying the graph-structure patterns, and better embeddings are selected to make the model perform classification tasks well.

\begin{table}[h]
\caption{Comparison of various classification indicators of ablation experiments.}
\label{tab1}
\centering
\begin{tabular}{p{35mm}<{\centering}p{13mm}p{13mm}p{13mm}p{13mm}}
\toprule
\multirow{2}{*}{Method} & \multicolumn{4}{c}{Metrics}                                       \\ \cmidrule{2-5}
                        & ACC            & PREC           & SEN            & SPEC           \\ \midrule
GSAN                    & 70.42          & 79.87          & 74.38          & 62.50          \\\specialrule{0em}{1pt}{1pt}
FGSAN(GAT-encoder)      & 77.92          & 84.97          & 81.25          & 71.25          \\\specialrule{0em}{1pt}{1pt}
FGSAN                   & \textbf{82.08} & \textbf{87.74} & \textbf{84.38} & \textbf{74.25} \\ \bottomrule
\end{tabular}
\end{table}

\begin{figure}[htbp]
	\centering
	\includegraphics[width=0.8\textwidth]{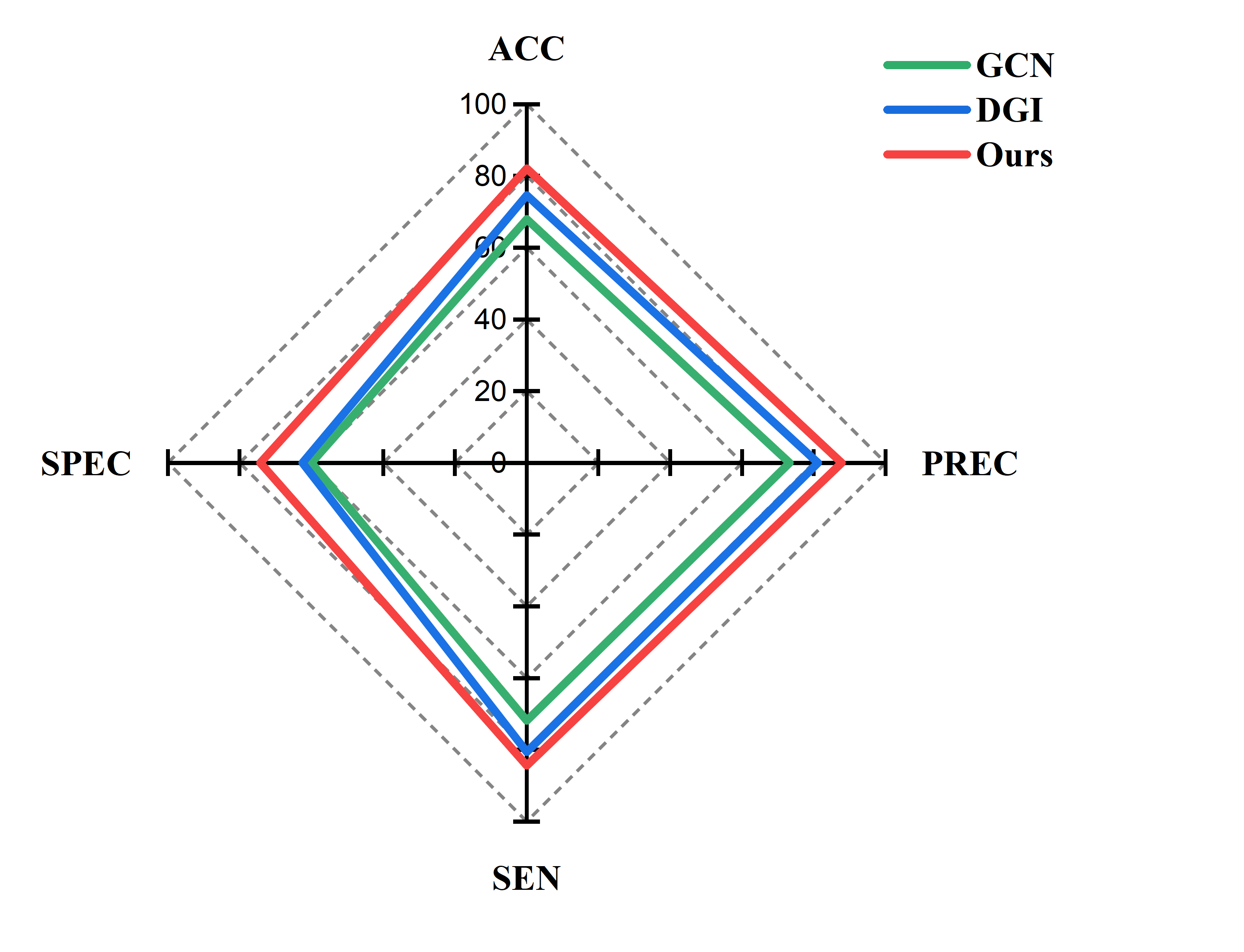}
	\caption{The performance comparison of FGSAN with other models. The method proposed in this paper is compared with the existing methods for classification experiments. The red line represents the method proposed in this paper.} \label{fig2}
\end{figure}

\begin{table}
	\caption{TOP five regional brain biomarkers extracted by the FGSAN model.}
	\label{tab2}
	\centering
	\begin{tabular}{c p{55mm}<{\centering}}
		\toprule
		No. & ROI name of biomarkers  \\ \midrule
		1   & Midbrain.R              \\\specialrule{0em}{1pt}{1pt}
		2   & Diagonal domain.R              \\\specialrule{0em}{1pt}{1pt}
		3   & Primary motor cortex.R  \\\specialrule{0em}{1pt}{1pt}
		4   & Hippocampal formation.L \\\specialrule{0em}{1pt}{1pt}
		5   & Insular cortex.L        \\ \bottomrule
	\end{tabular}
\end{table}
\subsection{Identification Performance}

This section conducts relevant comparative experiments to verify the superiority of FGSAN. The method proposed in this paper is compared with the existing GCN and DGI\cite{D2velickovic2019deep} methods. DGI learns node embeddings in an unsupervised manner. DGI can continuously optimize model results by maximizing the degree of correlation between two random variables. After multiple experimental verifications, it is found that our proposed FGSAN method is significantly better than existing methods in classification indicators. As shown in Figure \ref{fig2}, the method proposed in this paper has noticeable performance improvement compared with DGI and GCN. FGSAN outperforms the other two methods in every index of the binary classification experiment. The binary classification performance is outstanding on SPEC metrics, and FGSAN significantly outperforms the other two methods.

\begin{figure}
	\centering
	\includegraphics[width=0.9\textwidth]{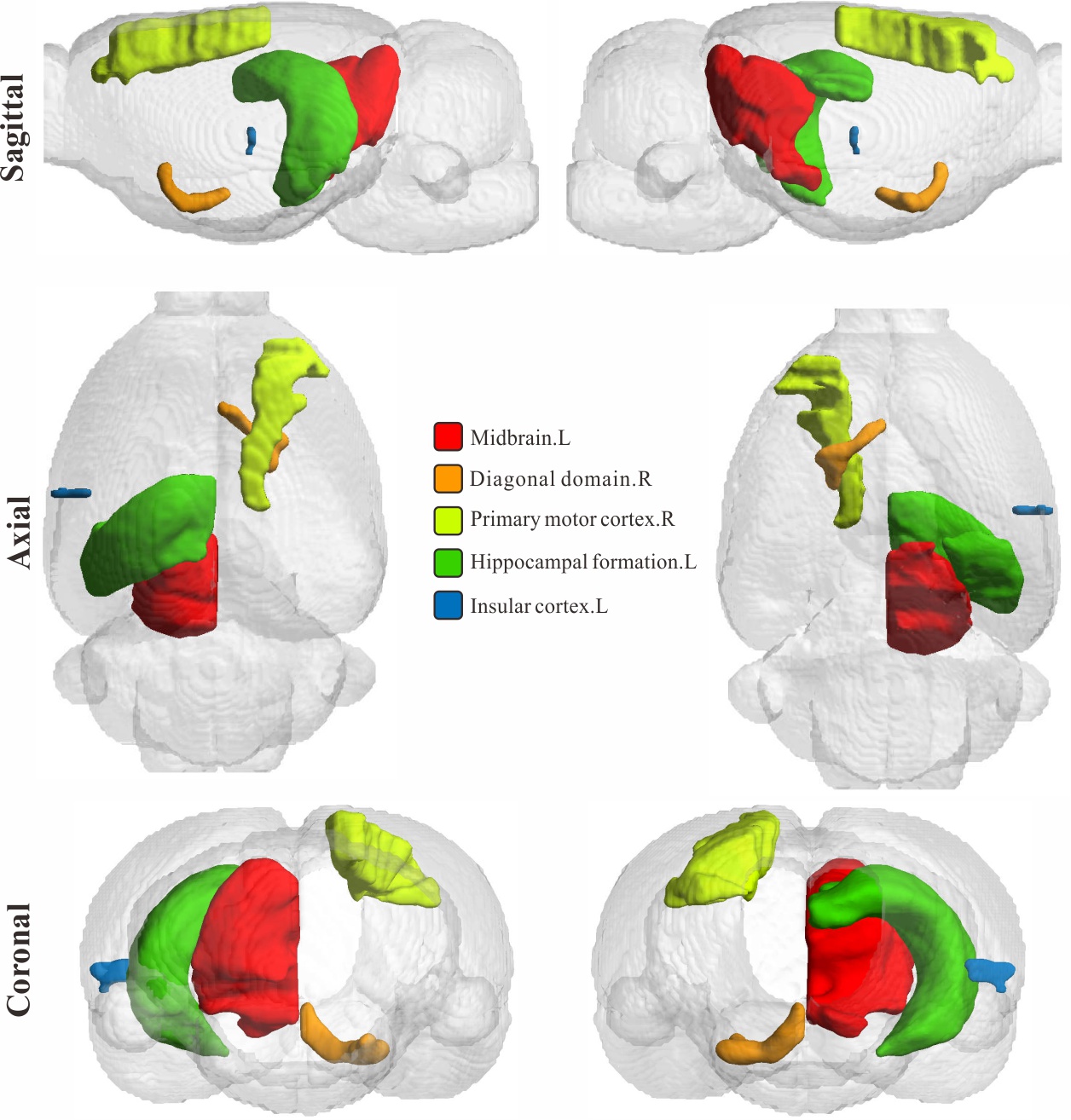}
	\caption{Visualization of top-five addiction-related brain regions. These five brain regions are all brain regions with higher weights output by the model. Their relevance to nicotine addiction was separately validated in previous work.} \label{fig3}
\end{figure}

\subsection{Interpretable Brain Regional Biomarkers}
The method presented here identified five brain regions with higher weights associated with nicotine addiction. As shown in Table 2, the five brain regions with higher weights are Midbrain.R\cite{16nguyen2021nicotine}, Diagonal domain.R\cite{17flannery2019habenular}, Primary motor cortex.R\cite{18smolka2006severity}, Hippocampal formation.I\cite{19ghasemzadeh2020expression}, and Insular cortex.L\cite{20keeley2020intrinsic}. These five brain regions have been proven to be associated with nicotine addiction in previous research work. The brain regions discovered by our model are recognized by experts in the relevant brain regions. We visualized the locations of these five brain regions. As shown in Figure 3, the locations of the five addiction-related brain regions found by the model in the rat brain are shown in three different directions.

\section{Conclusion}
In this research, we propose a new model called feature-selected graph spatial attention network(FGSAN) for exploiting effective and interpretable regional brain biomarkers and utilizing features of these biomarkers to identify the addiction-related brain network patterns. Detailed model discussions were conducted to examine the proposed FGSAN and the encoder's and feature selector's superiority, among other concerns. We obtained better results than the comparison method by using the selected graph representations for classification, indicating an advantage in graph feature extraction that may yield better graph embeddings in the latent space. And more significantly, these embeddings can be well explained in the neuroscience of addiction. Continuing our investigation into the brain processes of nicotine addiction in rats will focus on our future research.

\end{document}